\documentclass[12pt, a4paper]{article}
\usepackage{amsmath, amssymb}
\usepackage[french]{babel}
\usepackage[dvips]{epsfig}
\begin{document}
\title{\large\bf R\'eponse \`a un \'echelon de champ d'un supraconducteur de 
type II:\\
un moyen simple de tester l'ancrage des vortex en volume.\\
Magnetic-field step response of a type-II superconductor as a simple 
test of the vortex bulk pinning.}
\author{
        \normalsize
        H. Vasseur, P. Mathieu, B. Pla\c{c}ais et Y. Simon\\ 
        \footnotesize Laboratoire de Physique de la Mati\`ere 
        Condens\'ee de l'\'Ecole Normale Sup\'erieure,\\ 
        \footnotesize 24, rue Lhomond, F-75231 Paris Cedex 5}

\maketitle

Short title : step response of a type-II superconductor


\newpage
\begin{abstract}
Une lame en champ perpendiculaire est soumise \`a un \'echelon de 
champ magn\'etique de quelques Gauss d'amplitude et de m\^eme sens 
que le champ principal. La lame a \'et\'e d\'ecoup\'ee par \'etincelage ou 
lamin\'ee, sans pr\'ecaution particuli\`ere, \`a partir d'un \'echantillon de 
supraconducteur conventionnel (alliage de plomb-indium), et ses 
courants critiques ont des valeurs tout \`a fait standard.
 La mesure du champ \'electrique transitoire, 
induit sur une face de la lame, nous renseigne sur la p\'en\'etration
 du flux magn\'etique et l'entr\'ee des vortex par cette face. Le 
 seul \'ecart observ\'e \`a la r\'eponse id\'eale est que la diffusion 
 du champ magn\'etique est limit\'ee par un \'ecrantage superficiel, 
 associ\'e \`a une densit\'e de courant critique de surface. Sinon, le 
 r\'eseau de vortex dans la lame r\'epond exactement comme s'il n'y 
 avait aucun effet d'ancrage en volume.
 
 A slab in parallel field is submitted to a step-like magnetic-field 
 excitation, of amplitude 1--10 G, in the same direction as the 
 applied field. The slab has been rolled or spark cut, 
 without special care, from a conventional superconductor ingot 
 (lead-indium alloy), and its critical currents have standard values. 
 The transitory electric field, induced on a face of 
 the slab, has been measured. Thus, we obtain information about the 
 magnetic flux penetrating and vortices entering the sample through 
 this face. The only observed deviation from the ideal response is 
 that magnetic-field diffusion being limited by surface screening, 
 associated with a superficial critical-current density. Otherwise, 
 the vortex array inside the slab responds exactly as if bulk pinning 
 were ineffective.  
\end{abstract}

\newpage

\section{Introduction}

Une lame supraconductrice de type II, plong\'ee dans un champ 
magn\'etique ext\'erieur $\boldsymbol{B}_0$, est travers\'ee par un 
r\'eseau r\'egulier de lignes de vortex, sauf dans l'\'etat Meissner 
en champ parall\`ele faible ($B_0<B_{c1}$, le premier champ critique); 
c'est l'\'etat dit mixte (mixed state) g\'en\'eralement trait\'e comme 
un continuum \`a une \'echelle m\'esoscopique, grande devant la 
distance intervortex $a\sim$100--1000 \AA. Au dessus du second champ 
critique $B_{c2}$, et dans tous les cas de figure, la structure du 
r\'eseau de vortex dispara\^\i t dans la masse, o\`u le m\'etal 
redevient normal avec une r\'esistivit\'e $\rho_n$.

Si on ignore de petits effets li\'es aux d\'eformations de la maille 
du r\'eseau de vortex, par exemple les effets de cisaillement 
d\'ecrits par une tr\`es faible constante de cisaillement \'elastique
$c_{66}$, un \'etat du r\'eseau est bien d\'ecrit si on se donne en 
chaque point le champ de vortex $\boldsymbol{\omega} = 
n\varphi_0\boldsymbol{\nu}$, qui regroupe la densit\'e de lignes de vortex 
$n$ (m de lignes $/$ m$^3$) et leur direction $\boldsymbol{\nu}$ ($\nu=1$);
en multipliant par $\varphi_0$, quantum de flux, on donne 
arbitrairement \`a $\boldsymbol{\omega}$ la dimension d'un champ 
magn\'etique (en Teslas) \cite{Mathieu88,Hocquet92}. Lorsque 
l'\'echantillon est soumis \`a une excitation \'electromagn\'etique 
quelconque, champ magn\'etique variable ou courant appliqu\'e, le 
r\'eseau de vortex peut se mettre en mouvement, mouvement continu 
("flux flow") ou petites oscillations. Ces mouvements s'accompagnent 
toujours de dissipation et d'un champ \'electrique (m\'esoscopique) 
$\boldsymbol{E} = \boldsymbol{\omega}\times{\bf v}_L$, o\`u ${\bf v}_L$ est la vitesse de 
ligne  \cite{Mathieu88,Hocquet92}. 

Quel que soit le cas de figure, la r\'eponse \'electromagn\'etique de 
l'\'etat mixte met en jeu la dynamique des vortex. Un \'echantillon 
{\em id\'eal} n'ayant aucun d\'efaut cristallin, ni en volume ni en 
surface, donc aucun site d'ancrage (ou ``pinning'') possible pour les 
vortex, se comporterait assez banalement comme un milieu  conducteur et 
diamagn\'etique. Sa r\'esistivit\'e est anisotrope et varie de 
z\'ero (pour des courants parall\`eles aux vortex) \`a 
$\rho_f\simeq\rho_n\omega/B_{c2}$ (pour des courants normaux aux 
vortex). On d\'efinit d'autre part une ``perm\'eabilit\'e 
diamagn\'etique'' effective $\mu(\omega)<\mu_0$ \cite{Vasseur97}; 
la perm\'eabilit\'e relative $\mu_r=\mu/\mu_0$ est une fonction 
rapidement croissante du champ, de 0 \`a 1, si bien qu'aux champs $B$ 
assez \'elev\'es o\`u nous travaillons, $\mu$ se confond pratiquement 
avec $\mu_0$. 

\begin{figure}
 \centerline{\epsfig{file=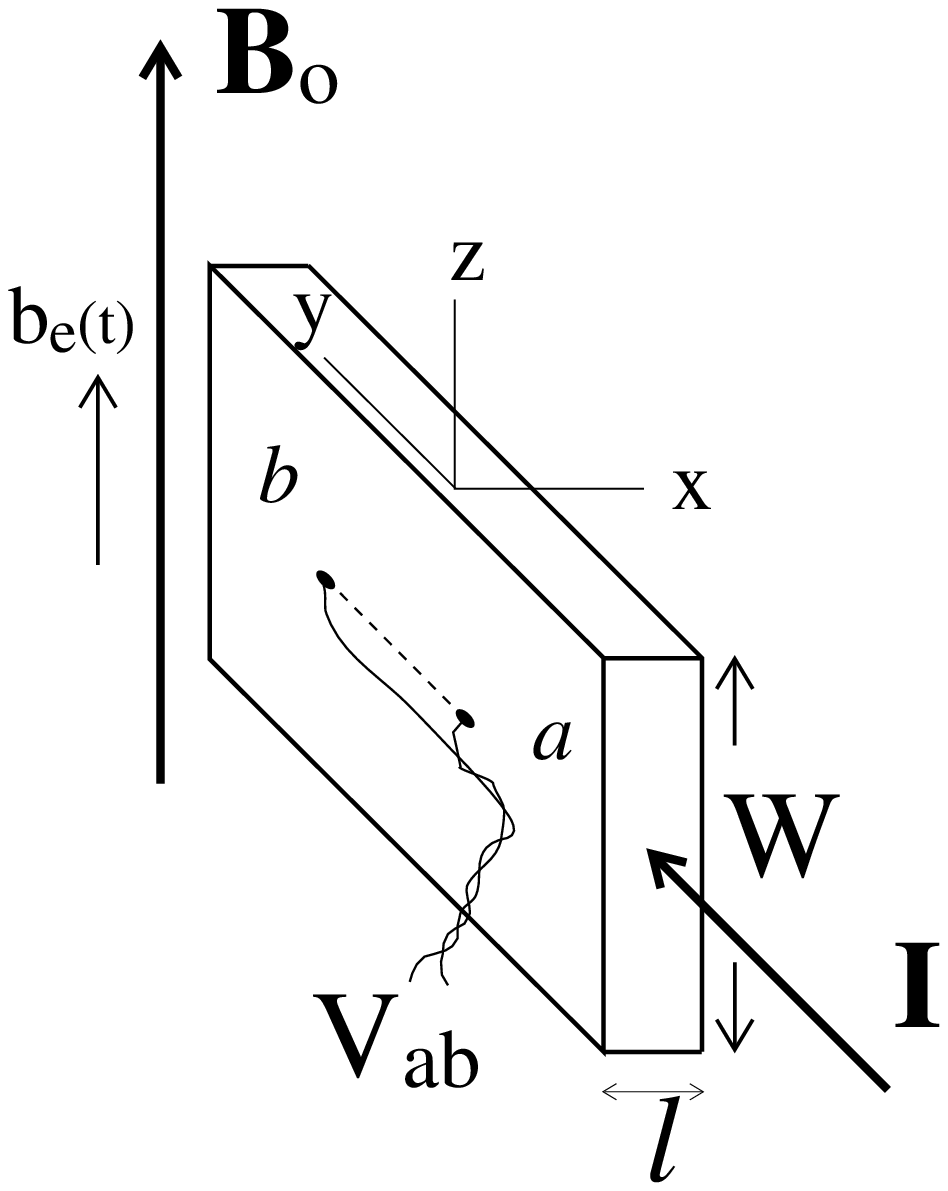, scale=1}}
\caption{\footnotesize Une lame de plomb-indium est immerg\'ee dans un 
champ magn\'etique variable ${\bf B}_0+{\bf b}_e(t)$ 
parall\`ele \`a ses faces. ${\bf B}_0 \geq 1000$ G est le 
champ principal appliqu\'e, et ${\bf b}_e(t)$ est une petite 
excitation p\'eriodique en cr\'eneaux ($\sim 100$ Hz), dont 
l'amplitude est de quelques Gauss. A chaque \'echelon croissant ou 
d\'ecroissant on observe une diffusion transitoire ($\sim100\;\mu$s) 
et  plus ou moins \'ecrant\'ee de la variation de champ ext\'erieur. 
Les courants $j(x,t)$ et champs \'electriques $e(x,t)$ induits sont 
dans la direction $y$. Deux prises de tension $a$ et $b$ sur la face 
$x=0$, s\'epar\'es de $d=5$--$10$ mm, permettent d'apr\'ecier la 
tension induite $e(0,t)d$, donc le flux magn\'etique ou encore le 
nombre de vortex entrant par la face $x=0$ entre $a$ et $b$. Un 
courant continu $I$ appliqu\'e dans la direction $y$ modifie 
consid\'erablement la r\'eponse (voir 
Fig.~\ref{reponse3}c).}
\label{slab1}
\end{figure}

Ainsi, la r\'eponse d'une lame \emph{id\'eale} \`a une 
excitation ext\'erieure ${\bf b}_e(t)$ dans la g\'eom\'etrie de la 
figure~\ref{slab1}, est solution d'une simple \'equation de diffusion
\begin{equation}
\frac{\partial^2b}{\partial x^2} = \mu_0\sigma_f\frac{\partial 
b}{\partial t}\qquad,
\label{diffusionequation}\end{equation}
o\`u $\sigma_f=\rho_f^{-1}$ est la conductivit\'e ``flux-flow''. Par 
exemple, la r\'eponse alternative id\'eale, pour un $b_0e^{-i\Omega 
t}$, serait un mode d'effet de peau classique d\'ecrit par 
l'\'equation de dispersion $k_1^2=i\mu_0\sigma_f\Omega=2i/\delta_f^2$ 
imm\'ediatement d\'eduite de (\ref{diffusionequation}), o\`u 
$\delta_f$ est la profondeur de peau ``flux-flow''. Ou encore, pour un 
courant continu $I$ appliqu\'e dans la direction $y$, la 
caract\'eristique courant-tension $V$-$I$ de la lame de la 
figure~\ref{slab1}, serait simplement une loi d'Ohm $V=R_fI$.

A cet \'egard, il est important de souligner que toutes les 
th\'eories sont d'accord sur la nature de cette r\'eponse id\'eale, 
mais aussi sur le fait qu'elle n'est jamais observ\'ee (sauf 
indirectement \cite{Vasseur97}), parce que les moindres d\'efauts 
cristallins affectent consid\'erablement la r\'eponse 
\'electromagn\'etique avec apparition de "pinning" et de courants 
critiques : la caract\'eristique $V$-$I$ en courant continu ne devient lin\'eaire 
qu'\`a fort courant : $V=R_f(I-I_c)$; la r\'eponse alternative de 
basse fr\'equence reste lin\'eaire \`a tr\`es faible niveau 
($b_0\lesssim1\mu$T), mais la profondeur apparente de p\'en\'etration est 
beaucoup plus faible que $\delta_f$, et de plus ind\'ependante de la 
fr\'equence \cite{Campbell69}, etc \dots 

En revanche, il existe de fortes divergences d'interpr\'etation sur 
la nature exacte des courants critiques et du processus de "pinning", de 
m\^eme que sur la localisation des pi\`eges (surface ou volume).

Nous pr\'esentons dans cet article une mesure de la r\'eponse \`a une 
petite variation en \'echelon du champ ext\'erieur, $\boldsymbol{B}_0+{\bf 
b}_e(t)$, d'une lame polycristalline mais chimiquement homog\`ene. 
Cette exp\'erience d\'emontre de fa\c{c}on tr\`es directe et 
spectaculaire, que dans ce type d'\'echantillon, dit ``soft'' (la 
distinction entre \'echantillons  ``soft'' et ``hard'' est 
effectivement tr\`es importante; nous en discuterons au
\S~\ref{modeles}), \emph{le 
r\'eseau de vortex en volume r\'epond  librement \`a une 
excitation}. Autrement dit, il n'y a pas de signe d\'ecelable de 
"pinning" en volume, et seuls les d\'efauts de surface gouvernent la 
r\'eponse globale. Ce r\'esultat corrobore le mod\`ele de l'\'etat 
critique propos\'e en 1988 par Mathieu et Simon (mod\`ele MS) 
\cite{Mathieu88}, en contradiction avec les id\'ees classiques sur la 
nature du "pinning" qui furent \'elabor\'ees il y a une trentaine 
d'ann\'ees.

La question d\'epasse en fait celle de la simple distribution des 
centres de "pinning", en surface ou en volume, et concerne tout autant 
la nature m\^eme du "pinning". Pour montrer l'importance de l'enjeu, 
nous pr\'esentons au \S~\ref{modeles}, aussi qualitativement que 
possible, un bref rappel des deux points de vue en pr\'esence, le point 
de vue classique, que nous appellerons en abr\'eg\'e BPM (``bulk pinning 
model'') d'une part, et le mod\`ele MS d'autre part. Nous pr\'ecisons 
\'egalement au \S~\ref{modeles} ce que nous entendons par 
mat\'eriau ``soft''. Le principe de l'exp\'erience et les r\'esultats 
sont pr\'esent\'es au \S~\ref{reponse}.

De nombreuses exp\'eriences  pr\'etendent mesurer des 
densit\'es volumiques de courant critique $J_c$ (A$/$cm$^2$), et 
semblent attester l'existence de "pinning" en volume dans toutes sortes 
d'\'echantillons.
En fait, bien souvent, l'existence de $J_c$ a \'et\'e postul\'ee, et 
le r\'esultat exp\'erimental est interpr\'et\'e via un mod\`ele o\`u 
$J_c$ est un param\`etre ajustable. L'exemple le plus banal est celui 
o\`u $J_c$ est tout simplement calcul\'ee comme le rapport d'un 
courant critique $I_c$ \`a la section de l'\'echantillon. Ou alors, si 
l'exp\'erience est cens\'ee mesurer v\'eritablement une distribution 
$J_c$ fonction du point, sans exclure {\it a priori} que $J_c$ puisse 
\^etre nulle en volume, une interpr\'etation sommaire conduit 
facilement \`a la conclusion erron\'ee que les $J_c$ de volume sont 
significatifs.

Ainsi van de Klundert {\it et al.} \cite{Klundert78}, \'etudiant la 
r\'eponse d'une lame (Fig.~\ref{slab1}) \`a une excitation $b_e(t)$ 
trap\'ezo\"\i dale, trouvent bien un courant critique de surface, mais 
aussi une distribution $J_c(x)$ en volume, dont la contribution \`a 
$I_c$ n'est pas du tout n\'egligeable. Mais curieusement cette 
distribution $J_c(x)$ pr\'esente un pic marqu\'e au centre $x=l/2$
de la lame; 
comme les auteurs le notent eux-m\^emes, cette anomalie de densit\'e 
de courant critique en $x=l/2$ est tout \`a fait 
invraisemblable. A notre avis, il s'agit l\`a d'un artefact de 
d\'epouillement d\^u \`a ce que la r\'eponse est consid\'er\'ee comme 
quasistatique, sous-estimant les effets de diffusion du champ 
magn\'etique (c.f. fin du \S~\ref{reponse}).

Citons \'egalement \`a ce propos les exp\'eriences plus r\'ecentes 
\cite{Abulafia95}, qui mettent en \'evidence dans des 
plaquettes en champ perpendiculaire, un profil du champ en tas de 
sable ("pile-of-sand profile"), conform\'ement au "critical state model"
de Bean \cite{Bean62} (CSM). Ces exp\'eriences sont souvent donn\'ees \`a 
tort comme des v\'erifications du bulk "pinning". Cette confusion entre 
CSM et BPM (``bulk pinning model'') est assez courante et nous y reviendrons au \S~\ref{modeles}.

\section{Les mod\`eles de "pinning"}\label{modeles}

Pour fixer les id\'ees, et sauf avis contraire, nous nous 
r\'ef\'ererons \`a la g\'eom\'etrie de la Fig.~\ref{slab1}, que nous 
avons utilis\'ee dans nos exp\'eriences : une lame $(xy)$ en champ 
$\boldsymbol{B}_0 (0,0,B_0)$ parall\`ele, dont les dimensions selon $x,y,z$ 
sont not\'ees respectivement $l$ (\'epaisseur), $L$ (longueur) et 
$W$ (largeur), avec $l\ll W<L$.

Rappelons le sch\'ema des th\'eories classiques. Notons d'abord que, 
sauf exception \cite{Coffey92}, elles confondent syst\'ematiquement 
le champ de vortex $\boldsymbol{\omega}$ et le champ magn\'etique $\boldsymbol{B}$ 
(moyenne locale) \cite{Mathieu88,Hocquet92}. Par contre elles 
distinguent artificiellement dans la densit\'e de courant $\boldsymbol{J}$, 
courant diamagn\'etique $\boldsymbol{J}_D$ et courant de transport 
$\boldsymbol{J}_T$ \cite{Campbell72}. En effet, comme le remarquait 
d\'ej\`a Josephson \cite{Josephson66}, un courant de transport 
non-dissipatif et un $\boldsymbol{J}_D$ sont tous deux des moyennes locales 
de m\^emes courants microscopiques ${\bf j}_s$. La force motrice qui 
tend \`a mettre les vortex en mouvement est la force de Lorentz 
$\boldsymbol{J}_T\times\varphi_0\boldsymbol{\nu}$ (par unit\'e de longueur de 
vortex), soit $\boldsymbol{J}_T\times\boldsymbol{\omega}=\boldsymbol{J}_T\times\boldsymbol{B}$ 
par unit\'e de volume. S'il n'y avait pas de "pinning", on aurait
$\boldsymbol{J}_T\times\varphi_0\boldsymbol{\nu}=\eta{\bf v}_L$, d'o\`u un 
flux-flow ${\bf v}_L$ \`a angle droit du courant de transport (le 
coefficient de friction $\eta$ est reli\'e \`a $\rho_f$ par 
$\eta=\varphi_0\omega/\rho_f$, et $\boldsymbol{E} = \rho_f\boldsymbol{J}_T$). Mais 
on pense que toutes sortes de d\'efauts cristallins peuvent se 
comporter comme des centres de "pinning". La force d'ancrage sur un 
vortex, dont le c\oe ur est voisin d'un site d'ancrage, est vue comme 
le gradient d'un profil d'\'energie libre fonction de la position du 
c\oe ur; elle se transmet \'eventuellement aux autres vortex par le 
jeu des interactions entre vortex. Ces interactions sont 
traditionnellement d\'ecrites, pour de petits \'ecarts \`a 
un r\'eseau id\'eal uniforme, par trois modules \'elastiques 
$c_{11}$ (compression), $c_{44}$ (torsion) et $c_{66}$ 
(cisaillement) \cite{Campbell72}. Les forces de "pinning", dont l'effet 
est en g\'en\'eral isotrope, sont suppos\'ees pouvoir compenser la 
force de Lorentz $\boldsymbol{J}_T\times\boldsymbol{B}$  jusqu'\`a une valeur 
seuil $J_T=J_c(B,T)$, dite densit\'e 
de courant critique, qui peut \^etre aussi fonction du point.
Dans ces conditions, le "flux flow" commencera dans une tranche d$y$ de 
la lame, quand partout sur cette section $J_T=J_c$; le courant 
critique $I_c(y)$, valeur du courant $I$ appliqu\'e, pour lequel on 
observe la premi\`ere tension aux bornes de cette tranche, est donc 
identifi\'e \`a la somme des $J_c$ sur la section.

Dans les interpr\'etations classiques du "pinning" et des courants 
critiques, il arrive qu'on envisage une forte contribution de la 
surface, dans le sens o\`u une forte densit\'e de pi\`eges peut se 
trouver localis\'ee sur la surface, ou du moins sur une faible 
profondeur (quelques microns), tout simplement \`a cause du traitement 
de surface ou de la technique de d\'ecoupe. Mais, que leur 
concentration soit faible ou \'elev\'ee, on n'envisage jamais 
l'absence de centres de "pinning" en volume; c'est pourquoi nous 
caract\'erisons les mod\`eles classiques par le terme de BPM. Si les centres d'ancrage sont trop dilu\'es en 
volume pour pouvoir pi\'eger les vortex un par un, on fait intervenir 
l'effet du module de cisaillement $c_{66}$ (pourtant tr\`es faible) 
\cite{Campbell72}, pour expliquer que l'ensemble 
du r\'eseau ne se met pas en mouvement imm\'ediatement au moindre 
courant appliqu\'e. C'est ainsi qu'on attribue la disparition de tout 
courant critique le long d'une ligne dite d'irr\'eversibilit\'e 
$B^*(T)$ dans les cuprates supraconducteurs, \`a une fusion du 
r\'eseau de vortex, qui ferait effectivement dispara\^\i tre tout 
effet de cisaillement. Dans d'autres cas, au contraire, on suppose 
que la concentration des centres d'ancrage en volume est importante, 
entra\^\i nant de nombreuses distorsions du r\'eseau de vortex en 
volume, au point d'en faire parfois un verre de vortex \cite{Blatter94}.

Avant d'\'enoncer quelques difficult\'es des BPM, rappelons ce qu'on 
observe lorsque le courant continu appliqu\'e \`a la lame est 
surcritique. Pour $I\!>\!I_c(y)$, la tension d$V$ aux bornes de la 
tranche d$y$ suit une loi lin\'eaire $dV\!=\!dR_f(I\!-\!I_c(y))$ 
o\`u $dR_f\!=\!\rho_fdy/Wl$. Comme il est pratiquement 
impossible d'obtenir une distribution parfaitement homog\`ene des 
d\'efauts, quel que soit le mod\`ele invoqu\'e, il est clair que 
$I_c(y)$ n'est pas constant sur la longueur $L$ de la lame, mais varie 
sur un intervalle $(I_c^\prime,I_c^{\prime\prime}$), de sorte que la caract\'eristique 
$V$-$I$ de la lame est une somme de caract\'eristiques 
\'el\'ementaires lin\'eaires, ce qui donne : $V=0$ jusqu'\`a $I_c^\prime$, 
caract\'eristique courb\'ee de $I_c^\prime$ \`a $I_c^{\prime\prime}$, et $V=R_f(I-I_c)$ 
au del\`a de $I_c^{\prime\prime}$, avec $I_c=\langle I_c(y)\rangle$, moyenne sur $L$.
Cette remarque pratique interviendra dans notre discussion du 
\S~\ref{reponse}. Rappelons \'egalement qu'\`a c\^ot\'e de cette 
tension continue en "flux flow", appara\^\i t une tension bruyante 
$\delta V(t)$ (typiquement $10^{-8}$--$10^{-11}$ V$/$(Hz)$^{1/2}$, 
dans la gamme 0--10 kHz), dont tout le monde s'accorde \`a dire 
qu'elle r\'esulte du mouvement plus ou moins irr\'egulier des vortex 
au voisinage des d\'efauts.

Une analyse des th\'eories classiques et la confrontation avec 
l'exp\'erience montrent assez vite que le BPM est un mod\`ele trop 
rustique, et qu'il est impuissant \`a rendre compte de fa\c{c}on 
coh\'erente de l'ensemble des r\'esultats exp\'erimentaux. 
Nous disons bien de l'ensemble des r\'esultats exp\'erimentaux.
Rappelons en effet que ce d\'ebat sur la nature et la localisation des 
courants critiques est, en ce qui nous concerne, commenc\'e depuis 
longtemps. Des exp\'eriences vari\'ees 
\cite{Hocquet92,Placais94,Mathieu93,LuetkeEntrup97}, dont certaines tr\`es 
anciennes \cite{Thorel72b,Thorel73}, remettent en cause la notion de 
densit\'e de courant critique et les m\'ecanismes classiques de "pinning" 
dans les ph\'enom\`enes de transport.
Il n'est \'evidemment pas question de reprendre ici en d\'etail tous les 
r\'esultats et arguments que nous avons accumul\'es sur le sujet; mais il 
est clair que nous avons d'ores et d\'ej\`a assez de r\'esultats pour 
affirmer que le BPM est r\'efut\'e dans toute une classe d'\'echantillons 
"soft", qu'il s'agisse d'alliages classiques, de m\'etaux purs, ou de 
cristaux d'YBCO non macl\'es ("untwinned crystals" voir plus loin).
Cependant, nous sommes bien conscients qu'on ne remet pas
aussi ais\'ement en question des notions qui datent de plus de 
trente ans. C'est la raison pour laquelle nous proposons ici une exp\'erience,
qui n'est jamais qu'une exp\'erience de plus, mais qui a la vertu d'\^etre 
tr\`es simple et d\'emonstrative, et surtout de pouvoir \^etre interpr\'et\'ee
ind\'ependamment de tout formalisme li\'e \`a un mod\`ele quelconque de 
"pinning".

Reprenons bri\`evement quatre exemples :\\ 
{\bf i)} Dans une lame inclin\'ee sur $\boldsymbol{B}_0$, 
dans un cylindre, une sph\`ere, il existe des densit\'es de courant 
diamagn\'etiques $\boldsymbol{J}_D$ tr\`es \'elev\'ees pr\`es de la 
surface ($\sim10^7$--$10^8$ A$/$cm$^2$); 
pourquoi ces courants ne donnent-ils lieu \`a aucune force de Lorentz ? \\
{\bf ii)} Si on compare les $\boldsymbol{J}_c$ (mesur\'ees comme les 
rapports $I_c/Wl$) obtenues avec des films, plaquettes, lames d'un 
m\^eme mat\'eriau et de diff\'erentes \'epaisseurs $l$, pr\'epar\'ees 
dans les m\^emes conditions \cite{Joiner67,Simon94}, on s'aper\c{c}oit que le plus 
souvent $J_c\propto 1/l$, ce qui signifie en clair que $I_c$ est 
proportionnel au p\'erim\`etre $2W$, ou ne varie pas si seulement $l$ 
varie. D'o\`u l'id\'ee qu'on ferait mieux de d\'efinir une densit\'e 
de courant critique superficielle, comme $K_c$(A$/$m)$=I_c/2W$. Cette 
id\'ee est confirm\'ee par des exp\'eriences faites dans notre 
laboratoire, et qui permettent de localiser directement $J_T$ ainsi 
que l'effet Joule \cite{Hocquet92}.\\
{\bf iii)} La forme de la caract\'eristique $V$-$I$ suppose que la 
force de "pinning" agit comme une force de frottement solide quand le 
r\'eseau de vortex est en mouvement; les 
mod\`eles classiques parviennent non sans mal \`a expliquer cette 
circonstance \cite{Campbell72}, mais ne sont jamais parvenus, sauf au 
prix de graves incoh\'erences \cite{Placais94}, \`a rendre compte du bruit de 
"flux flow". L'exp\'erience montre qu'il y a bien des fluctuations de 
vitesse de ligne $\delta v_L(t)$ dans la masse de l'\'echantillon, 
mais que, contrairement \`a toute attente, et en compl\`ete 
contradiction avec un BPM, elles sont coh\'erentes dans tout 
l'\'echantillon \cite{Placais94}.\\
{\bf iv)} tout le monde reconnait que la p\'en\'etration d'une onde 
\'electromagn\'etique dans une lame de type II, mesur\'ee par l'imp\'edance 
de surface $Z(\Omega)$ est gouvern\'ee par la dynamique du "pinning". 
Mais, comme nous l'avons montr\'e r\'ecemment 
\cite{LuetkeEntrup97,LuetkeEntrup98},
aucun BPM n'est capable d'expliquer tous les aspects qualitatifs de 
l'effet de peau dans l'\'etat mixte (effets de taille), et encore moins 
de rendre compte quantitativement du spectre $Z(\Omega)$ lorsqu'on 
parcourt la gamme des radiofr\'equences ("depinning transition") \cite{LuetkeEntrup98}.

Il nous para\^\i t d'autre part symptomatique qu'aucun BPM n'a \'et\'e 
capable de pr\'edire seulement l'ordre de grandeur observ\'e des 
courants critiques. Or le mod\`ele MS, en ne consid\'erant comme 
d\'efauts que la rugosit\'e de la surface, peut non seulement 
pr\'edire  cet ordre de grandeur \cite{Mathieu88,Hocquet92,Simon94}, 
mais apporte \'egalement une solution aux probl\`emes que nous avons 
\'evoqu\'es : caract\`ere superficiel des $I_c$, localisation en 
surface de la partie $VI_c$ de l'effet Joule $VI$, origine du bruit de 
"flux flow", existence d'une ligne d'irr\'eversibilit\'e dans les 
supraconducteurs anisotropes \cite{Simon94}, forme du spectre 
$Z(\Omega)$ dans des \'echantillons aussi vari\'es que PbIn, Nb, V, YBCO.

Nous renvoyons \'egalement aux articles originaux pour ce qui concerne les d\'etails 
du mod\`ele MS des courants critiques, et de la th\'eorie 
ph\'enom\'enologique dont ce mod\`ele d\'ecoule 
\cite{Mathieu88,Hocquet92,Simon94}. Nous nous contenterons  de 
donner ici trois \'el\'ements essentiels de la th\'eorie MS, qui, \`a 
notre avis, donnent la cl\'e de tous les probl\`emes de transport dans 
les supraconducteurs de type II : \\
{\bf i)} Chaque ligne de vortex doit 
se terminer perpendiculairement \`a la surface de l'\'echantillon, 
d'o\`u l'importance des conditions aux limites en surface (lisse ou 
rugueuse) dans tout probl\`eme d'\'equilibre ou de mouvement du 
r\'eseau de vortex. \\
{\bf ii)} Les lignes de vortex ne sont pas 
toujours les lignes de champ, de sorte que $\boldsymbol{\omega}$ et 
$\boldsymbol{B}$ doivent \^etre consid\'er\'es comme deux variables locales 
ind\'ependantes. La variable conjugu\'ee de $\boldsymbol{\omega}$, 
$\boldsymbol{\varepsilon}=\varepsilon(\omega,T)\boldsymbol{\nu}$, se pr\'esente 
comme une tension de ligne $\varphi_0\varepsilon$ (J/m) dans 
l'\'equation MS d'\'equilibre des vortex (ou de non-dissipation 
$\boldsymbol{J}_s\!+\!\mbox{curl}\boldsymbol{\varepsilon}\!=\!0$). La distinction entre 
$\boldsymbol{\omega}$ et $\boldsymbol{B}$ introduit des degr\'es de libert\'e 
suppl\'ementaires, et conduit \`a des solutions d'\'equilibre 
inattendues (d\'ecrivant des \'etats sous-critiques), qui ont 
\'echapp\'e aux th\'eories classiques.\\
 {\bf iii)} L'analogie 
classique du diamagn\'etisme local est trompeuse. Un courant 
diamagn\'etique $\boldsymbol{J}_D$ est un vrai courant supraconducteur 
non-dissipatif $\boldsymbol{J}_s$ ($=-\mbox{curl}\boldsymbol{\varepsilon}$), au 
m\^eme titre qu'un $\boldsymbol{J}_T$ sous-critique. L'un ou l'autre 
circule pr\`es de la surface, sur une faible profondeur $\lambda_V$
($\lesssim\lambda_0\sim1000$ \AA,  profondeur de p\'en\'etration de 
London \`a champ faible). Au del\`a de cette profondeur $\lambda_V$, 
tout \'ecart $\boldsymbol{\omega\!-\!B}$ dispara\^\i t, si bien que dans la 
masse $\boldsymbol{\omega}\!\equiv\!\boldsymbol{B}$. Il se trouve que l'intensit\'e 
d'aimantation moyenne d'un corps parfait dans l'\'etat mixte est 
justement $-\boldsymbol{\varepsilon}$, mais $-\boldsymbol{\varepsilon}$ n'a pas le sens physique 
premier d'une intensit\'e d'aimantation locale, ni $\mu_r$, d\'efini 
comme le rapport $\omega/(\omega+\mu_0\varepsilon)$ \cite{Vasseur97}, 
celui d'une v\'eritable perm\'eabilit\'e.

Pr\'ecisons maintenant la distinction que nous faisons entre 
\'echantillons "soft" et "hard" \cite{Mathieu88}.
La plupart des exp\'eriences fondamentales destin\'ees \`a explorer 
les propri\'et\'es de transport des supraconducteurs de type II, qu'il 
s'agisse de mat\'eriaux conventionnels ou de mat\'eriaux \`a haute 
$T_c$, utilisent des \'echantillons relativement homog\`enes 
chimiquement, que nous appelons ``soft''. Un \'echantillon ``soft'' 
peut \^etre un monocristal, ou une feuille polycristalline lamin\'ee 
pleine de d\'efauts cristallins, mais ses caract\'eristiques 
thermodynamiques telles que $B_{c1}, B_{c2}, \varepsilon$ ou $T_c$, 
sont bien d\'etermin\'ees et homog\`enes, ou du moins varient 
lentement \`a l'\'echelle m\'esoscopique de la distance intervortex $a$. 
Cette d\'efinition exclut bien entendu
tous les \'echantillons, dits ``hard'' (fils industriels, poudres 
fritt\'ees, \dots), qui contiennent des cavit\'es, pr\'ecipit\'es, 
d\'efauts colomnaires, \dots introduisant dans la masse de 
v\'eritables \emph{interfaces} \`a l'\'echelle de $a$. 
Dans ce sens, un cristal anisotrope macl\'e ("twinned crystal")
doit \^etre class\'e parmi les \'echantillons "hard";  un plan de 
macle ("twin boundary") repr\'esente une v\'eritable interface \`a 
l'\'echelle de $a$, et de belles exp\'eriences de STM \cite{MaggioAprile97}
montrent que cette interface peut transporter de fortes densit\'es de courant,
$J\sim 10^8$ A/cm$^2$, tout a fait comparables \`a celles que peut 
transporter une surface rugueuse dans le mod\`ele MS de l'\'etat critique.

La pr\'esence d'interfaces revient \`a multiplier artificiellement les effets de 
surface mis en \'evidence dans les \'echantillons ``soft''. On se 
souvient par exemple que les premiers fils supraconducteurs 
fabriqu\'es en France, avaient des courants critiques proportionnels 
\`a leur p\'erim\`etre, ou \`a leur rayon \cite{Thorel72}, tandis 
que les cables multi-filaments actuels ont des $I_c$ variant comme 
leur section. De m\^eme qu'en optique, o\`u il est pr\'ef\'erable de 
commencer par la physique du dioptre plut\^ot que par celle du tas de 
billes ou du verre cath\'edrale, il nous semble que la 
compr\'ehension du "pinning" et des courants critiques dans les 
\'echantillons ``hard'' (certes les plus utiles) passe d'abord par 
l'\'etude des ``soft''. Ce sont les seuls qui nous concernent ici.

Revenons sur un point improtant de la discussion \'evoqu\'e \`a la fin de 
l'introduction. La v\'erification du CSM dans des lames en champ normal 
est souvent pr\'esent\'ee comme une \'evidence exp\'erimentale de l'ancrage 
des vortex en volume, comme si CSM impliquait BPM.
Or l'id\'ee du CSM de Bean \cite{Bean62}, qui est ind\'ependante de la 
nature et de la localisation des courants critiques,
et peut en fait s'appliquer \`a tout mod\`ele des courants critiques, est 
simplement la suivante : la p\'en\'etration du champ (par exemple croissant) 
est limit\'ee par la saturation progressive, jusqu'\`a leur valeur critique et dans le sens des 
courants induits, des courants non dissipatifs, de l'ext\'erieur vers l'int\'erieur, ou pour 
une lame de la p\'eriph\'erie vers le centre. Notre analyse de la r\'eponse 
\`a un \'echelon (\S~\ref{reponsesymetrique}) est tout \`a fait conforme 
au CSM. R\'ecemment une \'equipe isra\'elienne propose un nouveau CSM dans 
des films qui tient compte de la variation de la densit\'e de courant 
critique sur l'\'epaisseur des films \cite{Prozorov98}.
Insistons sur le fait que si la technique \'el\'egante des petites sondes 
de  Hall permet bien de se rendre compte de la distribution des 
courants dans le plan de la lame, elle ne permet pas en revanche de 
r\'esoudre le probl\`eme de la distribution des courants sur l'\'epaisseur 
de la lame \cite{Abulafia95}.

\section{R\'eponse \`a un \'echelon}\label{reponse} 

\subsection{\'Echantillon et principe de l'exp\'erience}\label{sample}

Les exp\'eriences ont \'et\'e r\'ealis\'ees sur une s\'erie de lames 
polycristallines de plomb-indium. Cet alliage, dont les 
propri\'et\'es supraconductrices sont bien connues \cite{Farrel69},
a l'avantage de pouvoir \^etre pr\'epar\'e en lingots de grande 
taille ($\lesssim 10$ mm de diam\`etre). Le m\'elange de PbIn est 
fondu \`a 360$^\circ$C pendant plusieurs heures sous une pression de 
$3\times10^{-4}$ mbar d'Argon, puis tremp\'e \`a la temp\'erature 
ambiante. Un recuit progressif de 15 jours \`a une temp\'erature 
inf\'erieure de 8$^\circ$C \`a celle du point de fusion assure une 
bonne homog\'en\'eit\'e chimique de la solution solide 
Pb$_{1-x}$In$_x$ (pour $x\lesssim0.2$). Nous estimons que cette 
homog\'en\'eit\'e est satisfaisante si la largeur de la transition 
$\Delta B_{c2}$ n'exc\`ede pas 50 Gauss (soit $\Delta B_{c2}\lesssim0.01 B_{c2}$), 
ce que nous pouvons v\'erifier par une mesure de tension transverse an champ 
parall\`ele mise au point dans notre laboratoire \cite{Mathieu93}.
Les lames sont obtenues par \'electro-\'erosion, suivie ou non d'un 
laminage ou d'une compression entre plaques de verre. Les 
\'echantillons sont chimiquement homog\`enes, mais insistons sur le 
fait qu'aucune pr\'ecaution sp\'eciale n'a \'et\'e prise pour 
\'eviter les d\'efauts cristallins en volume, ou pour r\'eduire les 
courants critiques; ceux-ci sont d'un ordre de grandeur tout \`a fait 
standard.

Pour fixer les id\'ees, nous donnerons ci-dessous les valeurs 
num\'eriques et r\'esultats explicites obtenus avec une m\^eme lame 
Pb$_{0.82}$In$_{0.18}$, de dimensions $l=2.7$ mm, $W=7.2$ mm, et $L=30$ mm,
\`a $T=1.79$ K (soit 0.26 $T_c$). Son champ critique $B_{c2}$, \`a 
cette temp\'erature, est de 4750 G, et sa conductivit\'e normale est 
$\sigma_n=9.7\times10^6 \;\Omega^{-1}$m$^{-1}$.

La lame est soumise, dans la g\'eom\'etrie de la Fig.~\ref{slab1}, \`a 
une perturbation ${\bf b}_e(t)$ uniforme de m\^eme direction $z$ que 
le champ principal $\boldsymbol{B}_0$, et \'eventuellement \`a un courant 
continu $I$ superpos\'e dans la direction $y$ de la longueur de 
l'\'echantillon. Nous avons utilis\'e syst\'ematiquement une 
excitation $b_e(t)$ p\'eriodique en cr\'eneaux, d'amplitude $\pm 
b_0$ variable ($b_0\sim$ 1--10 G). La p\'eriode, de l'ordre de 
quelques ms, est grande devant le temps de diffusion du champ ($\sim 
100 \mu$s), de sorte qu'un \'etat d'\'equilibre est atteint \`a 
chaque demi-alternance. Le probl\`eme th\'eorique correspondant est 
celui de la r\'eponse  \`a un \'echelon, de $-b_0$ \`a $+b_0$ (ou de 
$+b_0$ \`a $-b_0$), et nous consid\'ererons qu'il peut \^etre 
trait\'e \`a {\em une dimension} suivant l'\'epaisseur de la lame 
$\Delta x = l\sim 1$ mm ($l\ll W,L$) : il s'agit donc dans chaque cas 
de calculer le profil $b_z=b(x,t)$ dans la lame, et d'en d\'eduire le 
champ \'electrique induit $e_y=e(x,t)$; la valeur du champ $e$ sur les 
faces de la lame est accessible \`a la mesure et peut \^etre 
compar\'ee \`a sa valeur th\'eorique. Pour fixer les id\'ees, nous 
consid\'ererons l'\'echelon croissant, en prenant comme origine des 
temps, $t=0$, l'instant o\`u le champ excitateur commence \`a cro\^\i 
tre ($b_e(t)=-b_0$ pour $t\leq0$).

A l'\'equilibre le champ principal $\boldsymbol{B}_1$ \`a l'int\'erieur de 
la lame est l\'eg\`erement plus faible que $\boldsymbol{B}_0$ \`a cause des 
courants diamagn\'etiques  superficiels. Le profil de champ $b(x,t)$ 
repr\'esente l'\'ecart par rapport \`a cet \'equilibre, et de m\^eme nous 
appelons $J$ ou $K$ toute densit\'e de courant induite par l'excitation 
ou associ\'ee \`a la pr\'esence d'un courant continu appliqu\'e $I$. Dans la 
discussion, comme sur les sch\'emas des figures \ref{reponse2} et 
\ref{reponse3},  il n'y aura pas d'inconv\'enient \`a ignorer les champs 
et courants d'\'equilibre. Comme nous envisageons la possibilit\'e de 
courants superficiels $K(0)$ et $K(l)$ sur les faces $x=0$ et $x=l$, 
dans la direction $y$ (\`a l'\'echelle de $l$ leur profondeur de 
p\'en\'etration $\lambda_V$ est negligeable),
 le champ $b$ a une discontinuit\'e $\mu_0K$ \`a 
chaque face, alors que $e_y$ est continu. Si on note $b(0,t)$ et 
$b(l,t)$ les valeurs du champ sur les faces mais \`a l'int\'erieur, 
on a 
\begin{eqnarray}
\mu_0K(0) &=& b_e(t) + b_I - b(0,t)\qquad ,\nonumber\\
\mu_0K(l) &=& b(l,t) - b_e(t) + b_I\qquad ,
\label{deux}\end{eqnarray}
o\`u $b_I=\mu_0I/2W$ est le champ \'eventuellement cr\'e\'e \`a 
l'ext\'erieur de la lame (c\^ot\'e $x<0$) par un courant $I$ appliqu\'e.

On mesure la tension induite $V_{ab}$ entre deux contacts $a$ et $b$, 
plac\'es \`a une distance $\Delta y=ab=d$ sur la face $x=0$ 
(Fig.~\ref{slab1}). Si $\Phi=sb_e(t)$ est le flux de $b_e(t)$ dans la 
boucle de mesure (de surface \'equivalente $s$) suppos\'ee ferm\'ee 
par le segment $ab$, $V_{ab}=\partial\Phi/\partial t+e(0,t)d$. Une 
fois retranch\'e le signal parasite $\partial\Phi/\partial t$, 
accessible \`a champ nul $B_0=0$, on obtient le signal utile $e(0,t) 
d$, qui mesure le flux rentrant par la face $x=0$ entre $a$ et $b$. 
Pour minimiser la surface de la boucle de mesure ($s<0.5$ mm$^2$), 
donc le signal parasite, les fils des prises de tension (diam\`etre 
5/100 mm) sont coll\'es sur la face de l'\'echantillon. Le signal 
transitoire $V_{ab}(t)$ ($\sim 100 \mu$V, voir Fig.~\ref{pulse4}), 
reproduit p\'eriodiquement \`a chaque mont\'ee de cr\'eneau, est 
amplifi\'e 1000 fois et analys\'e point par point par un int\'egrateur 
Boxcar PAR 160.

D'habitude, dans ce type d'exp\'erience le champ de surface est 
mesur\'e en enroulant une bobine autour de l'\'echantillon, avec 
l'avantage de multiplier le signal par le nombre de tours. Nous avons 
cependant pr\'ef\'er\'e la technique des prises de tension, quitte 
\`a amplifier le signal. Les m\^emes prises de tension permettent en 
effet de mesurer la tension continue associ\'ee \`a un courant 
$I$ appliqu\'e, et donc le courant critique $I_c$ (moyen) entre $a$ 
et $b$, qui est une donn\'ee essentielle; une bobine emp\^echerait 
l'acc\`es commode des prises de courant, et n'\'evite pas les effets de 
bout. D'autre part, une bobine prendrait en compte l'entr\'ee du flux 
par les deux faces de la lame; cela n'a pas d'importance si les faces 
jouent un r\^ole sym\'etrique et que $e(l,t) = -e(0,t)$; mais nous 
verrons qu'un courant $I$ appliqu\'e rompt cette sym\'etrie. Chaque 
vortex transportant un quantum de flux, la tension $V_{ab}$ mesure le 
nombre de vortex (par seconde) qui entrent dans la lame par la face 
$x=0$, et par cette face seulement.

\subsection{R\'eponse sym\'etrique $(I=0)$}\label{reponsesymetrique}

D\'ecrivons d'abord la r\'eponse {\em quasistatique}, c'est-\`a-dire 
la r\'eponse \`a une oscillation lente du champ $b_e(t)$ de $-b_0$ \`a 
$b_0$, qui donne les m\^emes profils limites du champ $b(x)$ 
qu'avant et apr\`es la mont\'ee d'un \'echelon (Fig.~\ref{reponse2}). 
Partant d'un \'etat d'\'equilibre, supposons que $b_e(t)$ diminue 
jusqu'\`a $-b_0$, puis remonte \`a $+b_0$. Si la surface a la 
capacit\'e de transporter un courant $K$ non-dissipatif, avec 
un maximum critique $K_c$, on s'attend \`a ce que l'\'ecrantage 
des variations du champ ext\'erieur soit parfait tant que 
$b_0<b_c=\mu_0K_c$. Notons en passant que cette conclusion est conforme 
au CSM.

Si $b_0$ d\'epasse $b_c$, on peut envisager deux cas de figure pour la 
r\'eponse quasistatique (Fig.~\ref{reponse2}). Ou 
bien, il n'y a pas de "pinning" en volume $(J_c=0)$, et l'exc\`es de 
champ $b_0-b_c$ p\'en\`etre librement et uniform\'ement dans la lame 
comme dans un m\'etal ordinaire (Fig.~\ref{reponse2}b); le champ 
int\'erieur oscille alors entre deux valeurs $\pm b_0^\prime$ avec 
$b_0^\prime=b_0-b_c$. Ou bien il y a du "pinning" en volume, disons avec 
un $J_c$ uniforme. Dans ce cas le profil de champ se complique 
(Fig.~\ref{reponse2}a). Conform\'ement au CSM, $J_y=-(1/\mu_0)\partial 
b/\partial x = \pm J_c$ ou 0. En cons\'equence pour des \'ecarts 
$b_0-b_c$ pas trop \'elev\'es, la p\'en\'etration du champ est 
limit\'ee \`a une profondeur $x_0<l/2$ :
\begin{equation} 
x_0 = \frac{b_0-b_c}{\mu_0J_c}\qquad .
\label{equation2}\end{equation}

Consid\'erons maintenant, dans la premi\`ere hypoth\`ese $(J_c=0)$, 
la r\'eponse \`a un \'echelon parfait d'amplitude $b_0>b_c$ (Fig.~\ref{reponse3}b). Des 
courants induits $+K_c$ et $-K_c$ s'\'etablissent 
imm\'ediatement sur les faces $x=0$ et $x=l$, imposant les conditions 
aux limites $b(0,t) = b(l,t) = b_0^\prime = b_0-b_c$. La r\'eponse 
$b(x,t)$ est la solution de l'\'equation de diffusion (\ref{diffusionequation}) 
satisfaisant ces conditions aux limites, avec le profil initial, \`a 
$t=0^+$ : $b(0,0)=b(l,0)=b_0^\prime$ et $b(x,0)\equiv-b_0^\prime$ 
ailleurs (Fig.~\ref{reponse3}b). Cette solution s'exprime 
analytiquement par d\'ecomposition en modes de Fourier pour $t\geq0$ :
\begin{equation}
b(x,t)=b_0^\prime\left[1-\sum_{\rm n \; impairs}\frac{8}{n\pi}\;\sin\left[\frac{n\pi 
x}{l}\right]\;\exp\left(-\frac{t}{\tau_n}\right)\right]\quad,\quad\mbox 
{o\`u}\quad\tau_n=\frac{\mu_0\sigma_fl^2}{n^2\pi^2}\quad;
\label{champb}\end{equation}
Le temps de diffusion est gouvern\'e par la constante de temps la plus 
longue, $\tau_1\simeq0.1\mu_0\sigma_fl^2$. La Fig.~\ref{reponse3}b 
montre sch\'ematiquement l'\'evolution du profil de $b$ dans la lame. 
La diffusion du champ bien entam\'ee \`a $t=\tau_1$ est pratiquement 
termin\'ee \`a $t=5\tau_1$ (profil plat $b_0^\prime$, limite 
$t=\infty$ de l'expression (\ref{champb}). On en d\'eduit le champ 
\'electrique induit sur les faces $x=0$ et $x=l$ par 
$e_y=j_y/\sigma_f=(1/\mu_0\sigma_f)\partial b/\partial x$, 
conform\'ement \`a l'\'equation de diffusion (\ref{diffusionequation}), puisque 
$-\partial b/\partial t=\partial e/\partial x$
\begin{equation}
e(0,t)=-e(l,t)=\frac{8b_0^\prime}{\mu_0\sigma_fl}
\sum_{\rm n \; 
impairs}\;\exp\left(-\frac{t}{\tau_n}\right)\quad.\qquad(t>0)
\label{champe}\end{equation}

Dans la seconde hypoth\`ese du BPM, avec $J_c=const.$ et un \'ecart 
$b_0-b_c$ pas trop \'elev\'e, la r\'eponse sch\'ematis\'ee sur la 
Fig.~\ref{reponse3}a est plus complexe; elle a \'et\'e calcul\'ee 
num\'eriquement par Kawashima {\it et al.} \cite{Kawashima78}. La 
diff\'erence essentielle avec la r\'eponse libre (\ref{champb}) est 
que le temps caract\'eristique de diffusion, 
$\tau\sim\mu_0\sigma_fx_0^2$ (voir la figure~9 de la 
Ref.~\cite{Kawashima78}), doit varier maintenant avec la profondeur de 
p\'en\'etration du champ $x_0$ (Eq.~\ref{deux}), donc avec l'amplitude 
de l'\'echelon. Le temps de diffusion devrait augmenter comme 
$(b_0-b_c)^2$.

Or nous n'observons rien de tel. A champ  $B_0$ assez \'elev\'e, nous 
trouvons effectivement un seuil $b_0=b_c$ au del\`a duquel il y a une 
p\'en\'etration massive du flux magn\'etique. En toute rigueur, 
l'\'ecrantage de l'excitation en dessous de $b_c$ n'est pas parfait, 
m\^eme aux plus faibles niveaux, \`a cause d'un faible effet de peau 
lin\'eaire bien connu \cite{Campbell69,Alais67}; ce ph\'enom\`ene, 
strictement, est en contradiction avec le CSM. Mais les signaux 
$e(0,t)$ associ\'es sont plus de 10 fois plus petits que notre signal 
parasite. Nous pouvons par cons\'equent les n\'egliger dans cette 
exp\'erience, o\`u le seuil $b_c$ reste en pratique bien marqu\'e. Le 
fait exp\'erimental essentiel est que la forme du signal transitoire 
$e(0,t)$ observ\'e pour de faibles \'ecarts $b_0-b_c$ reste la m\^eme 
quand $b_0$ augmente; un changement d'\'echelle permet de superposer 
tr\`es exactement les signaux obtenus pour diff\'erentes valeurs de 
$b_0>b_c$. Cela veut dire que la cin\'ematique de la diffusion est 
ind\'ependante de l'amplitude de l'excitation, en accord avec 
l'\'equation (\ref{champe}). Autrement dit il n'y a aucune variation 
d\'etectable du temps de diffusion, qu'elle soit quadratique en 
$b_0\!-\!b_c$ ou autre, comme le voudrait un BPM.

D'autres que nous ont mis en \'evidence un seuil $b_c$ 
\cite{Klundert78}, et par ailleurs la possibilit\'e d'une densit\'e de 
courant critique superficielle $K_c$ n'est pas contest\'ee. Mais
curieusement, aucun auteur n'a cherch\'e \`a comparer la 
contribution $2WK_c$ de la surface au courant critique total 
$I_c=2WK_c+WJ_cl$, mesur\'e directement et ind\'ependamment \`a 
partir d'une caract\'eristique continue courant-tension. Or notre 
montage permet de mesurer \`a la fois $I_c$ et $b_c$, et nous avons 
toujours trouv\'e que $I_c\simeq2WK_c=2Wb_c/\mu_0$, 
c'est-\`a-dire qu'\`a la pr\'ecison des mesures $J_c=0$. Ainsi, \`a 
4000 G, nous mesurons d'une part $I_c=10.2$ A, et d'autre part 
$b_c=9.0\pm 0.1$ G, ce qui correspond \`a une densit\'e de courant 
critique $K_c=b_c/\mu_0$ voisine de 7 A/cm, et un courant 
critique $2WK_c=10.3\pm0.1$ A.

Plus quantitativement on peut songer \`a comparer la tension 
transitoire induite, $e(0,t)\;d$, \`a la solution th\'eorique 
(\ref{champe}) de l'\'equation de diffusion. Mais, dans notre cas, le 
temps de mont\'ee de $b_e(t)$ est de l'ordre de 10 $\mu$s; cela se 
voit directement sur la dur\'ee de l'impulsion de tension parasite 
$s\;db_e/dt$ (Fig.~\ref{pulse4}). Ce temps de mont\'ee 
\'etant comparable au temps de diffusion $\tau_1$ lui-m\^eme, il est 
clair que le calcul (\ref{champb}) doit \^etre corrig\'e, si on 
esp\`ere un accord quantitatif.

Si $t=0$ est toujours l'instant o\`u $b_e(t)$ commence \`a cro\^\i 
tre, la p\'en\'etration du champ est retard\'ee et commence seulement 
\`a l'instant $t_d$, o\`u $b_e(t_d)$ atteint la valeur $-b_0^\prime + 
b_c=-b_0+2b_c$, apr\`es quoi $b(0,t)\!=\!b(l,t)$ cro\^\i t 
progressivement de $-b_0^\prime $ \`a $+b_0^\prime$. La condition 
limite de l'\'echelon fini id\'eal $b(0,t)\!=\!b_0^\prime\left[{\rm 
Y}(t)\!-\!1\right]$ est formellement remplac\'ee par une somme infinie 
d'\'echelons infinit\'esimaux d\'ecal\'es dans le temps; la r\'eponse 
s'obtient alors par superposition :
\begin{equation}
e(0,t)=-e(l,t)=\frac{4}{\mu_0\sigma_fl}
\sum_{\rm n \;impairs}\;\int_{{t}_d}^t 
{\frac{db_e}{dt^\prime}\;\exp\left(-\frac{t^\prime-t}{\tau_n}\right)}\quad{\rm 
d}t^\prime\quad,\qquad(t>t_d)
\label{champebis}\end{equation}
tandis que $e(0,t<t_d)=0$. Noter qu'avec $t_d=0$ et 
d$b_e/$d$t^\prime=2b_0^\prime\delta(t^\prime)$ dans (\ref{champebis}) 
on retrouve bien l'\'equation (\ref{champe}) pour un \'echelon id\'eal.

Il se trouve que la forme du signal parasite $s\;$d$b_e/$d$t$, 
ind\'ependante de l'\'echantillon et de l'amplitude du cr\'eneau, est 
assez bien repr\'esent\'ee par la diff\'erence de deux exponentielles 
$A(e^{-t/\theta_1}-e^{-t/\theta_2})$, en prenant $\theta_1=1.6\;\mu$s 
et $\theta_2=0.8 \;\mu$s. D'une telle expression (purement empirique) on 
d\'eduit une expression de d$b_e/$d$t$ (et de $b_e(t)$), qui est un 
interm\'ediaire commode pour calculer la r\'eponse th\'eorique 
(\ref{champebis}) dans tous les cas de figure $(B_0,l,b_0$ et $b_c)$ :
\begin{equation}
\frac{db_e}{dt} = \frac{2b_0}{\theta_1-\theta_2} 
\;\left(\exp\left[-\frac{t}{\theta_1}\right]-\exp\left[-\frac{t}{\theta_2}\right]\right)\quad.
\label{phi}\end{equation}

Nous trouvons que la r\'eponse th\'eorique ainsi corrig\'ee d\'ecrit 
le signal mesur\'e de fa\c{c}on tr\`es satisfaisante. La figure 
\ref{pulse4} concerne un exemple de r\'eponse asym\'etrique observ\'e 
en pr\'esence d'un courant continu, mais elle illustre bien la 
qualit\'e de l'accord quantitatif obtenu dans tous les cas de figure. 
Cet accord constant atteste que la diffusion du champ \`a 
l'int\'erieur de la lame, comme le mouvement des vortex qui 
l'accompagne, n'est ni limit\'e, ni g\^en\'e par des d\'efauts de volume.

\subsection{R\'eponse asym\'etrique $(I\neq0)$}\label{reponseasymetrique}

Pour confirmer notre analyse, nous avons calcul\'e et mesur\'e la 
r\'eponse \`a un \'echelon en pr\'esence d'un courant continu $I$ 
appliqu\'e. On introduit ainsi une dissym\'etrie entre les deux 
faces, qui conduit \`a un effet curieux sur le temps de diffusion.

La figure~\ref{reponse2}c montre la r\'eponse quasistatique dans le 
sc\'enario suivant. Partant toujours d'un \'equilibre, 
un courant de transport $I<I_c$ est appliqu\'e dans la 
direction $y$; le champ ext\'erieur devient $+b_I$ d'un c\^ot\'e de 
la lame ($x<0$) et $-b_I$ de l'autre. Chaque face transporte dans 
le m\^eme sens une nappe superficielle de courant de densit\'e $K<K_c \,(b_I<b_c)$. Puis le champ appliqu\'e est diminu\'e 
jusqu'\`a $-b_0$ : le courant induit sur la face $x=l$ est de 
m\^eme sens que le courant de transport de sorte que $K(l)$ 
augmente. Si l'amplitude de l'\'echelon est assez \'elev\'e pour que 
$b_I+b_0>b_c$, alors $K(l)$ atteint sa valeur de saturation 
$K_c$, tandis que la densit\'e de courant $K(0)$ sur 
l'autre face diminue et reste sous-critique. Dans ces conditions, les 
vortex vont p\'en\'etrer (et le champ diffuser) dans la lame par la 
face $x=l$ seulement, et le champ int\'erieur va diminuer 
jusqu'\`a $-(b_0+b_I-b_c)$. Ensuite le champ appliqu\'e remonte \`a 
$+b_0$ : les r\^oles des deux faces sont interchang\'es; les vortex 
entrent par la face $x=0$, et le champ int\'erieur plafonne \`a 
$+(b_0+b_I-b_c)$.

Consid\'erons maintenant la r\'eponse au cr\'eneau p\'eriodique, dans 
les m\^emes conditions d'amplitude : $b_I<b_c$, mais $b_I+b_0>b_c$. 
Le champ int\'erieur oscille entre deux profils plats 
$\pm(b_0+b_I-b_c)$. Le transitoire $e(0,t)$ \`a la mont\'ee d'un 
\'echelon (voir Fig.~\ref{reponse3}c) se calcule et se corrige comme 
plus haut, en rempla\c{c}ant simplement $b_c$ par $b_c-b_I$ (soit 
$b_0^\prime = b_0+b_I-b_c$), mais avec une importante diff\'erence 
li\'ee \`a la dissym\'etrie des conditions aux limites : en $x=0$, 
rien n'est chang\'e, $b(0,t)=b_e(t)+b_I-b_c$ suit la variation du 
champ appliqu\'e \`a partir de l'instant $t=t_d$, o\`u 
$b_e(t_d)=-b_0+2(b_c-b_I)$; mais en $x=l$ cette variation reste 
\'ecrant\'ee, et le fait que les vortex ne rentrent pas dans 
l'\'echantillon par cette face impose la nouvelle condition 
$e(l,t)\equiv0$ (ou $\partial b/\partial x=0$). Or cette situation est 
exactement celle de la r\'eponse sym\'etrique dans la moiti\'e d'une 
lame d'\'epaisseur $2l$; d'o\`u un effet spectaculaire, jamais 
signal\'e, se traduisant par un facteur 4 sur le temps de diffusion, 
puisque $\tau_1\propto l^2$.

Nous observons effectivement cet allongement du temps de diffusion, 
et mieux encore une co\"\i ncidence remarquable entre la r\'eponse 
calcul\'ee et le signal observ\'e (Fig.~\ref{pulse4}). Le bien-fond\'e 
de la condition $e\equiv0$ sur une face est confirm\'e directement par 
l'absence de tension induite sur la face $x=0$ aux demi-alternances 
correspondant \`a un \'echelon d\'ecroissant. La technique habituelle 
de mesure de l'effet de peau par une bobine captrice enroul\'ee sur 
l'\'echantillon ne permettrait pas de distinguer le r\^ole 
dissym\'etrique des deux faces. 

Terminons par quelques remarques sur les ordres de grandeur et les 
conditions pratiques, qui ont contribu\'e \`a rendre pr\'edictif un 
calcul relativement simple \`a partir d'une \'equation de diffusion 
1D. Prenons les donn\'ees de la figure~\ref{pulse4}. Nous avons 
d\'ej\`a indiqu\'e les dimensions de la lame ($l=2.7$ mm; $W=7.2$ 
mm); connaissant $\sigma_n$, la caract\'eristique $V$-$I$ dans 
l'\'etat normal donne la meilleure mesure de la distance $d=ab$ entre 
les prises de tension, soit $d=6.5$ mm. A 3000 G, la pente et 
l'abscisse \`a l'origine de la partie lin\'eaire de la 
caract\'eristique $V$-$I$ donnent respectivement la conductivit\'e 
flux-flow $\sigma_f=1.76\times10^7 \Omega^{-1}$m$^{-1}$, et le 
courant critique $I_c=20.4$ A. On en d\'eduit les constantes de temps 
de diffusion $\tau_1=\pi^{-2}\mu_o\sigma_f4l^2=65 \mu$s et 
$\tau_n=\tau_1/n^2$. Comme nous l'avons rappel\'e au \S~\ref{reponse}, 
le courant critique ainsi mesur\'e repr\'esente une valeur moyenne 
entre $a$ et $b$, $I_c(y)$ variant dans le cas pr\'esent entre 
$I_c^\prime\simeq20.1$ A et $I_c^{\prime\prime}\simeq20.7$ A. De 
m\^eme $K_c$ et $b_c=\mu_0I_c/2W\simeq17.8\pm0.3$ G. Une 
dispersion trop importante de $b_c$ pourrait remettre en cause le 
calcul \`a une dimension. Le point important ici n'est pas l'apparition 
de termes $\partial^2b/\partial y^2\sim d^2b_c/dy^2$ dans 
une \'equation de diffusion plus g\'en\'erale \`a 2D, car ces termes, 
comme on le v\'erifie ais\'ement, restent n\'egligeables devant 
$\partial^2b/\partial x^2$. En revanche des effets 2D significatifs 
peuvent r\'esulter du fait que, $t_d$ \'etant fonction de $y$, la 
diffusion en masse du champ ne commence pas partout en m\^eme temps le 
long de la lame. Nous pensons que c'est le cas dans les exp\'eriences 
de van de Klundert {\it et al.} \cite{Klundert78}, o\`u l'excitation 
$b_e(t)$ est trap\'ezo\"\i dale avec des rampes relativement lentes. 
Mais, pour les \'echelons rapidement croissants ou d\'ecroissants que 
nous utilisons, cet \'etalement de $t_d$ est n\'egligeable \`a 
l'\'echelle des temps de diffusion; ainsi dans le cas de figure 
\ref{pulse4} nous trouvons $t_d=1.6\pm0.4 \mu$s.

\section{Conclusion}\label{conclusion} 

La question de la localisation des courants et des centres d'ancrage, 
et plus fondamentalement la compr\'ehension de la nature m\^eme du 
"pinning", est \'evidemment essentielle pour la plupart des applications 
des supraconducteurs. Par ailleurs il est clair que l'efficacit\'e des 
d\'efauts de volume comme sites d'ancrage joue un r\^ole capital dans 
la physique actuellement tr\`es d\'evelopp\'ee des diff\'erentes 
phases du r\'eseau de vortex dans le diagramme $(B,T)$ des cuprates 
supraconducteurs.

Or, en d\'epit d'un nombre consid\'erable de travaux sur le sujet, 
d'abord sur des mat\'eriaux conventionnels dans les ann\'ees 1960 et 
le d\'ebut des ann\'ees 70, et plus r\'ecemment, apr\`es 1986, sur les 
nouveaux mat\'eriaux, nous pensons que le probl\`eme est loin d'\^etre 
r\'esolu. Un certain nombre d'exp\'eriences ant\'erieures 
\cite{Hocquet92,Placais94}, et d'autres plus r\'ecentes 
\cite{LuetkeEntrup97,LuetkeEntrup97} nous ont convaincu de l'inefficacit\'e de 
l'ancrage des vortex en volume dans toute une classe de mat\'eriaux ``soft'' 
telle que nous l'avons d\'efinie au \S~\ref{modeles}. 
L'exp\'erience que nous avons pr\'esent\'ee n'est qu'une exp\'erience de plus 
dont les conclusions  vont dans le m\^eme sens, et dont le seul 
int\'eret est la grande simplicit\'e. Elle ne suppose aucune 
th\'eorie pr\'eliminaire, en dehors des cons\'equences 
\'el\'ementaires des \'equations de Maxwell; son but est donc 
essentiellement p\'edagogique. Elle permet de v\'erifier asez directement 
le libre mouvement des vortex dans la masse. Soulignons que la 
g\'eom\'etrie d'une lame en champ parall\`ele est indispensable; l'analyse 
de la r\'eponse \`a une impulsion d'une lame en champ normal serait 
beaucoup plus compliqu\'ee.

Comme nous voulions simplement proposer un test exp\'erimental simple, 
nous n'avons pas cherch\'e ici \`a changer syst\'ematiquement d'\'echantillon.
La grande vari\'et\'e des \'echantillons (alliages, m\'etaux purs, YBCO, 
\ldots) auxquels notre mod\`ele de supraconducteur "soft" s'applique 
est d\'emontr\'ee et discut\'ee ailleurs \cite{LuetkeEntrup97,Simon94,LuetkeEntrup97}.
Terminons par deux remarques qui nuancent notre conclusion. 

Le probl\`eme n'est pas tant de montrer que la surface joue un r\^ole plus 
ou moins important dans le "pinning"; tout le monde est convaincu que 
les d\'efauts de surface y contribuent. Mais, en g\'en\'eral, on s'interroge 
simplement sur le poids relatif du volume et de la surface suivant le 
traitement de l'\'echantillon. Ce que nous pr\'etendons, c'est qu'il 
existe une vaste cat\'egorie d'\'echantillons dits "soft", faciles \`a 
obtenir, et pour lesquels tout se passe comme si (\`a la pr\'ecision des mesures 
pr\`es bien entendu) le r\'eseau de vortex r\'epondait librement dans la 
masse. Maintenant, il est aussi tr\`es facile d'introduire des 
interfaces dans la masse d'un \'echantillon et de le rendre "hard".

D'autre part, quand nous parlons de l'inefficacit\'e des sites 
d'ancrage en volume, il ne s'agit que des probl\`emes de transport.
Ce que nous constatons, c'est que les d\'efauts cristallins ne g\^enent 
pas le mouvement des vortex. Notre conclusion concerne donc seulement 
la dynamique des vortex (courants critiques, bruit de "flux flow", 
imp\'edance de surface, \ldots). Cela n'exclut pas que ces m\^emes 
d\'efauts cristallins dans la masse peuvent perturber la configuration 
du r\'eseau de vortex \`a l'\'equilibre, ce qu'on peut observer par 
diff\'erentes m\'ethodes d'imagerie (d\'ecoration ou diffraction de neutrons). 
Nous pensons notamment que tous les mod\`eles de "pinning collectif",  
qui relient le d\'esordre du r\'eseau de vortex aux distributions 
statistiques de points d'ancrage, sont sans doute pertinents, 
mais qu'il faut se garder d'\'etendre ces notions aux 
probl\`emes de dynamique des vortex et des courants critiques. 
C'est en tout cas ce que sugg\`ere l'exp\'erience.

\newpage
\addtocounter{figure}{-1}

\begin{figure}
\caption{\footnotesize 
  A lead-indium slab is immersed in a variable magnetic field ${\bf
    B}_0+{\bf b}_e(t)$ parallel to its faces. ${\bf B}_0 \geq 1000$ G
  is the applied principal field, and ${\bf b}_e(t)$ is a square
  waveform excitation ($\sim$ 100 Hz), a few Gauss in amplitude. Each
  increasing or decreasing step in the excitation gives rise to a
  transitory ($\sim$ 100 $\mu$s) and more or less screened diffusion
  of the variation of the external magnetic field. Induced currents
  $j(x,t)$ and electric fields $e(x,t)$ are in the $y$ direction.
  Voltage contacts $a$ and $b$ on the face $x=0$, a distance
  $d=5$--$10$ mm apart, allow the induced voltage $e(0,t)d$ to be
  measured, and, therefore, to estimate the magnetic flux, or else the
  number of vortices, entering through the face $x=0$ between $a$ and
  $b$.  Applying a dc current $I$ in the $y$ direction strongly alters
  the response, as shown in Fig.~\ref{reponse3}c.}
\end{figure}

\begin{figure}
\centerline{\epsfig{file=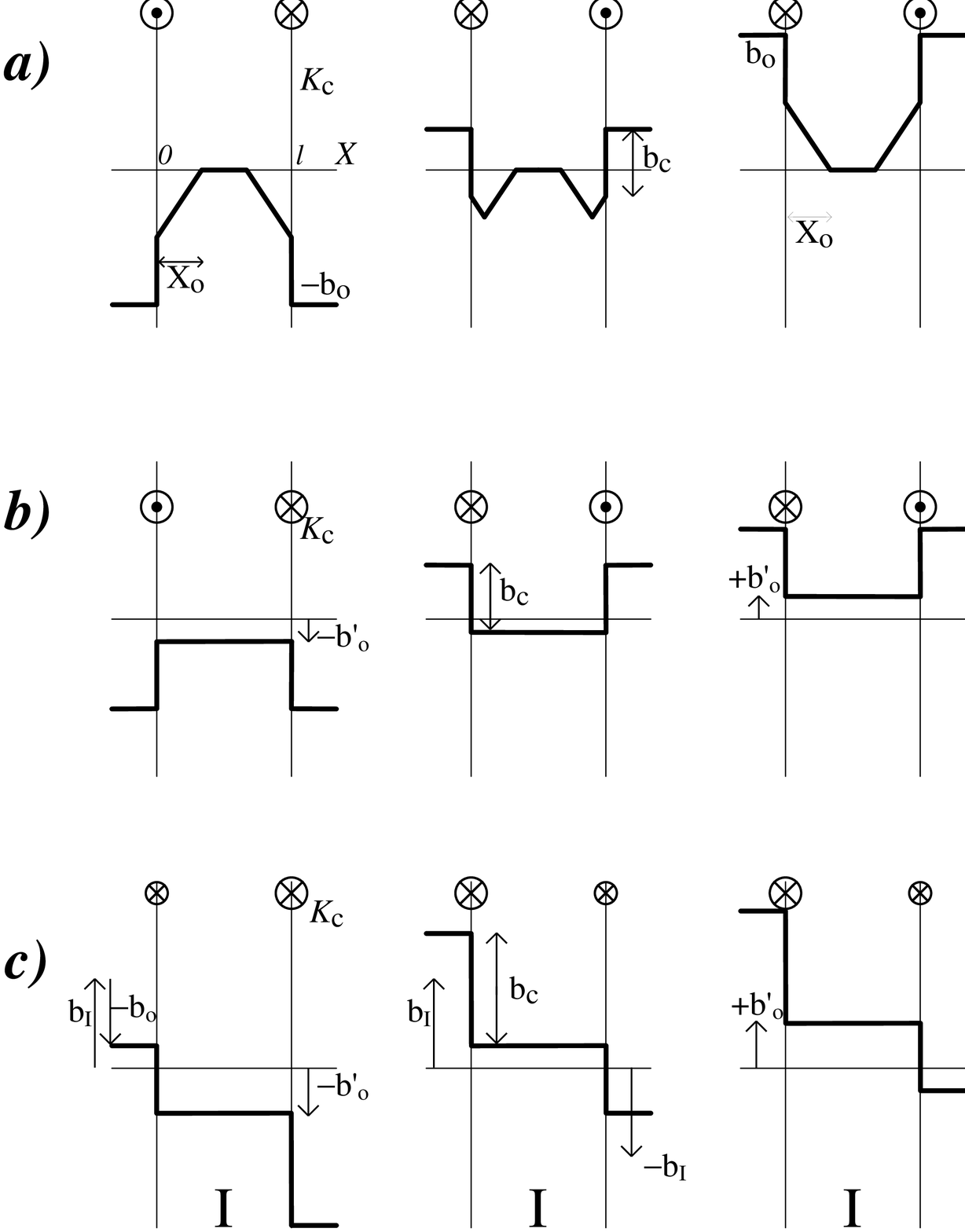, scale=0.4}}
\caption{\footnotesize 
  R\'eponse de la lame \`a une oscillation quasistatique du champ
  excitateur de $-{\bf b}_0$ \`a $+{\bf b}_0$, quand son amplitude
  $b_0$ d\'epasse le seuil critique d'\'ecrantage par les courants
  superficiels. Ce seuil not\'e $b_c=\mu_0K_c$ caract\'erise la
  capacit\'e de la surface \`a transporter un courant non dissipatif.
  Le trait \'epais repr\'esente le profil $b(x)$. {\bf a)} $b_0>b_c$,
  et "pinning" en volume caract\'eris\'e par une densit\'e de courant
  critique $J_c$ uniforme.  Pour les \'ecarts $b_0\!-\!b_c$ pas trop
  \'elev\'es la p\'en\'etration du champ est limit\'ee \`a une
  profondeur $x_0\propto b_0\!-\!b_c$. {\bf b)} $b_0>b_c$, et absence
  de "pinning" en volume; le champ int\'erieur est uniforme et oscille
  entre $-b_0^\prime$ et $b_0^\prime = b_0\!-\!b_c$. {\bf c)} Un
  courant continu $I$ est appliqu\'e, de fa\c{c}on que
  $b_I=\mu_0I/2W<b_c$, mais que $b_0\!+\!b_I>b_c$; il n'y a pas de
  "pinning" en volume. Le champ int\'erieur oscille entre
  $-b_0^\prime$ et $b_0^\prime = b_0\!+\!b_I\!-\!b_c$.}
\label{reponse2}\end{figure}
\addtocounter{figure}{-1}

\begin{figure}
\caption{\footnotesize 
  The response of the slab to a quasistatic oscillation of the
  exciting field, from $-{\bf b}_0$ to $+{\bf b}_0$, when its
  amplitude $b_0$ overcomes the critical threshold due to the
  screening effect of superficial currents.  This threshold, denoted
  as $b_c=\mu_0K_c$, characterizes the ability of the surface to carry
  a non-dissipative current. The thick line represents the field
  profile $b(x)$. {\bf a)} $b_0>b_c$, and bulk "pinning" characterized
  by a uniform critical current density $J_c$.  For small enough
  deviations $b_0\!-\!b_c$, the field penetration is limited to the
  depth $x_0\propto b_0\!-\!b_c$.  {\bf b)} $b_0>b_c$, and no bulk
  pinning; the internal field is uniform and oscillates between
  $-b_0^\prime$ and $b_0^\prime = b_0\!-\!b_c$.  {\bf c)} The slab is
  driven by a dc current $I$, so that $b_I=\mu_0I/2W<b_c$, but
  $b_0\!+\!b_I>b_c$; there is no bulk pinning.  The internal field
  oscillates between $-b_0^\prime$ and $b_0^\prime =
  b_0\!+\!b_I\!-\!b_c$.}
\end{figure}

\begin{figure}
\centerline{\epsfig{file=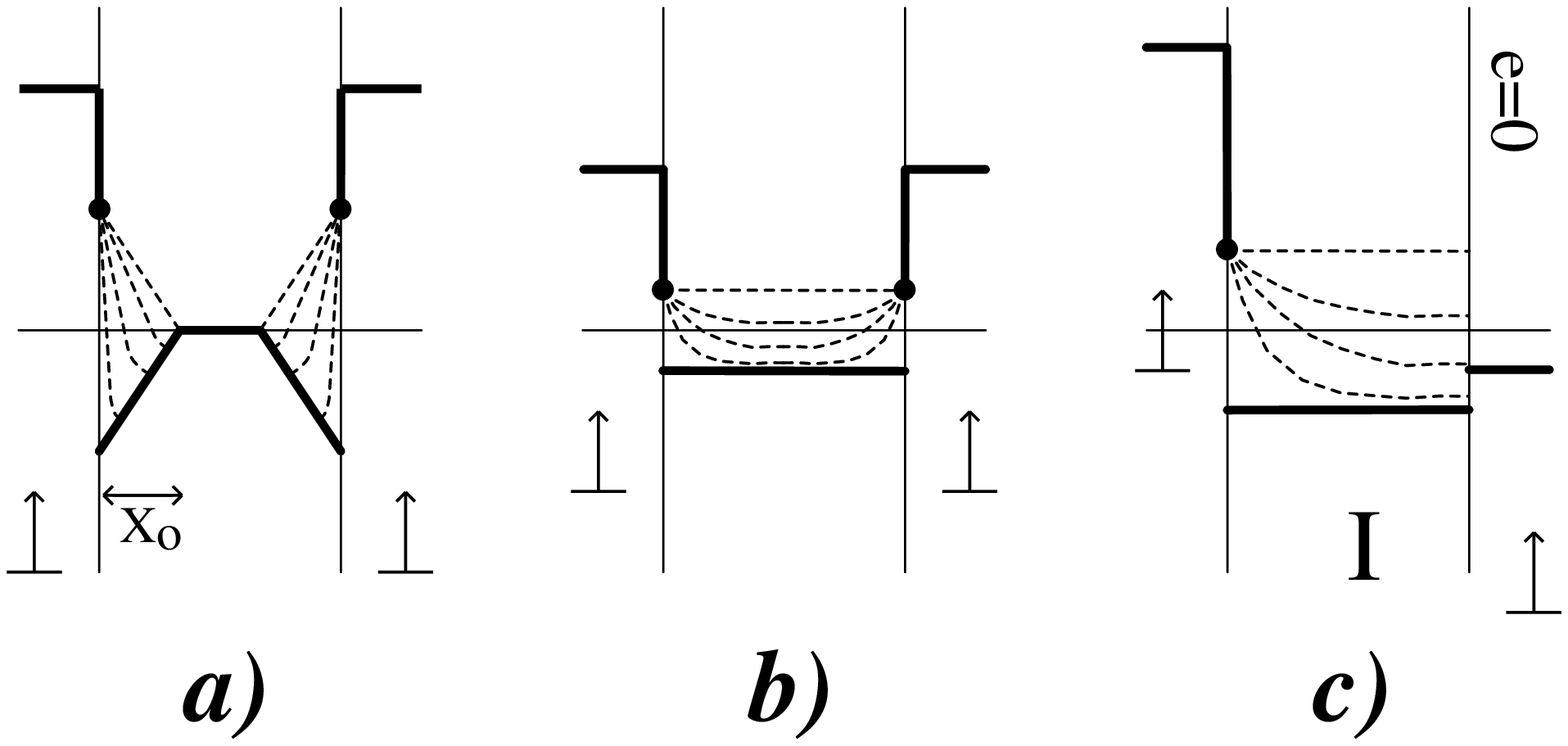, scale=0.6}}
\caption{\footnotesize 
  R\'eponse de la lame \`a un \'echelon id\'eal du champ appliqu\'e,
  dans les conditions a), b) et c) de la figure~{\ref{reponse2}}. Le
  trait \'epais repr\'esente le profil initial du champ \`a l'instant
  $t=0^+$, o\`u le champ appliqu\'e vient de passer brutalement de
  $-{\bf b}_0$ \`a $+{\bf b}_0$. Les courbes en tirets montrent
  l'\'evolution ult\'erieure du profil de champ $b(x,t)$ vers le
  profil d'\'equilibre (profil de droite Fig.~{\ref{reponse2}}); pour
  fixer les id\'ees, ce profil est repr\'esent\'e sch\'ematiquement
  \`a quatre instants $t=0.05 \tau$, $0.1 \tau, 0.5 \tau$ et
  $t=\infty$, o\`u $\tau=\mu_0\sigma_f X^2$ et $X$ sont un temps et
  une profondeur caract\'eristiques de diffusion qui d\'ependent du
  cas de figure : {\bf a)} La profondeur de p\'en\'etration $X=x_0$,
  donn\'ee par ({\ref{equation2}}), est limit\'ee par le "pinning" en
  volume; le temps de diffusion d\'epend donc de l'amplitude de
  l'excitation, ce qui n'est pas observ\'e exp\'erimentalement.
  L'\'evolution du profil a \'et\'e calcul\'ee en
  Ref.~{\cite{Kawashima78}}, en supposant une loi locale
  $J=J_c+\sigma_f e$. {\bf b)} La diffusion du champ est libre en
  volume, $X=l$ \'epaisseur de la lame, et $\tau$ est ind\'ependant de
  l'amplitude de l'\'echelon. Le profil $b(x,t)$ est la solution
  ({\ref{champb}}) de l'\'equation de diffusion
  ({\ref{diffusionequation}}).  {\bf c)} En pr\'esence d'un courant
  continu appliqu\'e, le champ ne diffuse, et les vortex ne
  p\'en\`etrent, que par la face de gauche $x=0$, tandis que le champ
  \'electrique $e$ est nul sur l'autre face. La r\'eponse
  asym\'etrique est la m\^eme que celle qu'on obtiendrait, pour le cas
  de figure b), dans la moiti\'e gauche d'une lame d'\'epaisseur $2l$.
  D'o\`u un facteur 4 sur $\tau$, qui est bien v\'erifi\'e
  exp\'erimentalement.}
\label{reponse3}\end{figure}
\addtocounter{figure}{-1}

\begin{figure}
\caption{\footnotesize 
  The response of the slab to an ideal steplike variation of the
  applied field, under the conditions a), b) and c) of
  figure~{\ref{reponse2}}.  The thick line represents the initial
  field profile at time $t=0^+$, where the field has increased
  suddenly from $-{\bf b}_0$ to $+{\bf b}_0$.  The dashed lines show
  the subsequent evolution of the field profile $b(x,t)$ towards the
  equilibrium profile (the right one in Fig.~{\ref{reponse2}}); for
  definiteness, this profile is sketched at four times denoted as
  $0.05 \tau$, $0.1 \tau, 0.5 \tau$ and $t=\infty$, where
  $\tau=\mu_0\sigma_fX^2$ and $X$ are characteristic diffusion time
  and depth depending on the situation : {\bf a)} The penetration
  depth $X=x_0$, as given by ({\ref{equation2}}), is restricted by the
  bulk pinning; the diffusion time thus depends on the amplitude of
  the excitation, at variance with experiment. The evolution of the
  field profile has been calculated in Ref.~{\cite{Kawashima78}},
  while assuming a local law $J=J_c+\sigma_fe$. {\bf b)} The field
  diffuses freely in the bulk, $X=l$ is the slab thickness, and $\tau$
  is independent of the step amplitude. The profile $b(x,t)$ is the
  solution ({\ref{champb}}) of the free diffusion equation
  ({\ref{diffusionequation}}). {\bf c)} If a dc current is applied,
  the field diffuses and vortices penetrate only through the left face
  $x=0$, while the electric field $e$ is zero on the other face. The
  asymmetric response is the same as would be obtained, in the case
  b), in the left half of a slab of thickness $2l$. Whence a factor 4
  in the diffusion time $\tau$, in agreement with experiment.}
\end{figure}

\begin{figure}
\centerline{\epsfig{file=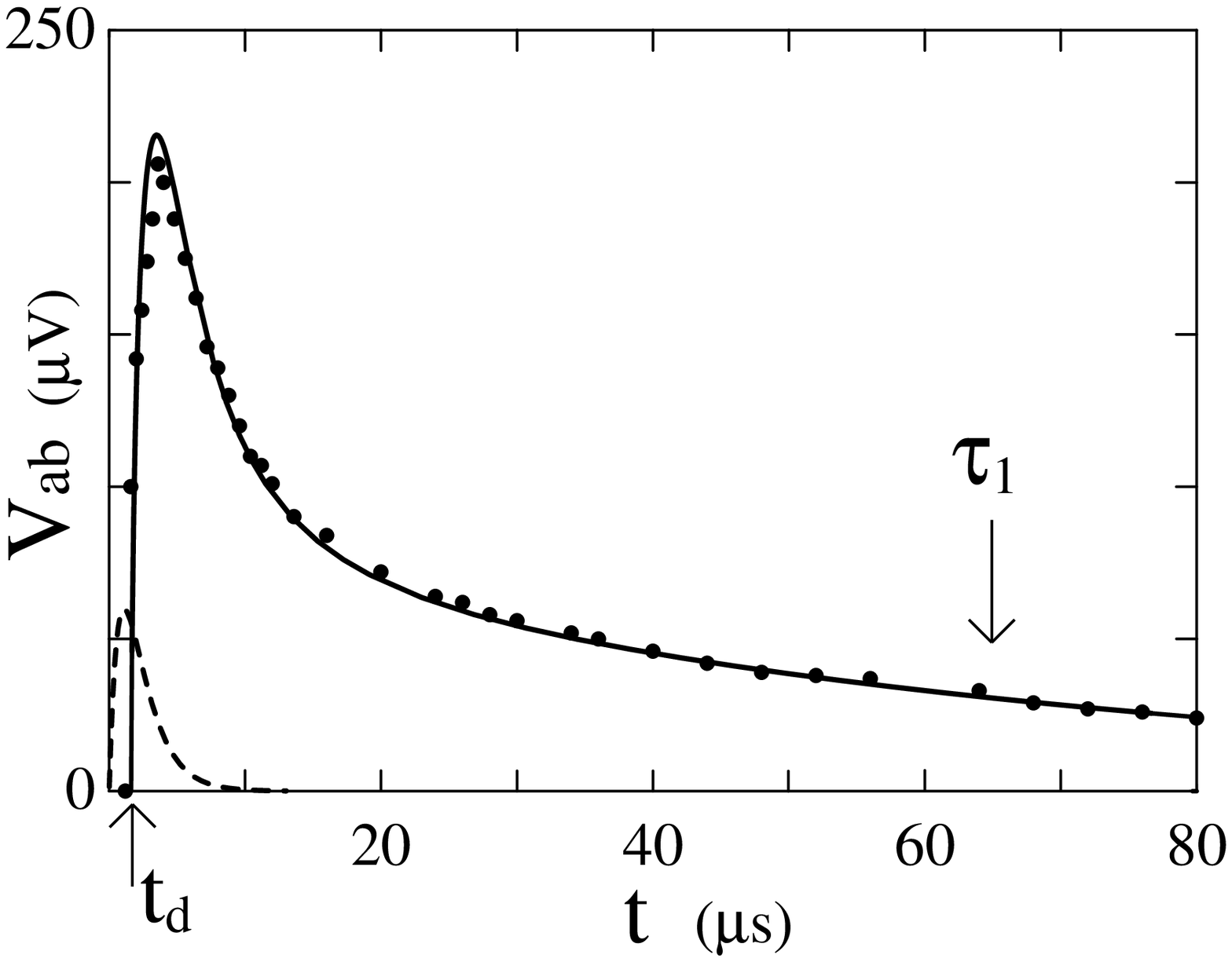, scale=0.45}}
\caption{\footnotesize 
  Tensions transitoires induites $V_{ab}(t)$ au moment de la mont\'ee
  d'un \'echelon $b_e(t)$ de $-b_0$ \`a $+b_0$ (voir
  Fig.~{\ref{slab1}}). L'\'echantillon est une lame polycristalline de
  Pb$_{0.82}$In$_{0.18}$, de dimensions $l\times L\times
  W=2.7\times30\times7.2$ mm$^3$. La distance $d$ entre les prises de
  tension est de 6.5 mm. La temp\'erature est 1.79 K.  Un courant
  continu $I=19$ A est appliqu\'e dans la direction $y$ ($b_I=16.6$
  G). L'amplitude de l'\'echelon est ici $2b_0=6$ G.  L'instant
  origine est pris au d\'ebut de la mont\'ee de l'\'echelon.  Le
  signal $V_{ab}(t)$ est amplifi\'e, puis analys\'e point par point
  par un int\'egrateur Boxcar. Le signal parasite induit dans
  la boucle de mesure (courbe en tirets) est mesur\'e \`a champ nul,
  $B_0=0$. Sa dur\'ee ($\sim10\;\mu$s) correspond au temps de mont\'ee
  de l'\'echelon. De sa forme, empiriquement bien repr\'esent\'ee par
  une diff\'erence de deux exponentielles, on d\'eduit $db_e/dt$ sous
  la forme ({\ref{phi}}), puis la forme $b_e(t)/b_0$ de l'\'echelon,
  qui est une constante instrumentale. L'amplitude du parasite fournit
  une estimation de la surface \'equivalente de la boucle de mesure,
  $s=0.32$ mm$^2$. Les donn\'ees de cette figure, obtenues en
  soustrayant le signal parasite, repr\'esentent les valeurs
  exp\'erimentales de la tension induite $e(t,0)d$, \`a 3000 G
  ($\bullet$). A 3000 G, $\sigma_f=1.76\times10^7\Omega^{-1}$m$^{-1}$,
  et $I_c=20.4$ A ($b_c=17.8$ G). Connaissant $b_c$ on d\'eduit que la
  diffusion commence th\'eoriquement avec le l\'eger retard
  $t_d=1.6\mu$s. La courbe en trait plein repr\'esente la r\'eponse
  th\'eorique $e(0,t) d$ obtenue en rempla\c{c}ant $l$ par $2l$ dans
  l'\'equation ({\ref{champebis}}).  La constante de temps
  $\tau_1\simeq0.1\tau=0.4\mu_0\sigma_fl^2\simeq65\mu$s est 4 fois
  plus longue qu'on attendrait dans une r\'eponse
  sym\'etrique. }
\label{pulse4}\end{figure}
\addtocounter{figure}{-1}

\begin{figure}
\caption{\footnotesize 
  Induced transitory voltages $V_{ab}(t)$ during an increasing step of
  $b_e(t)$ from $-b_0$ to $+b_0$ (see Fig.~{\ref{slab1}}). The sample
  is a Pb$_{0.82}$In$_{0.18}$ polycristalline slab with dimensions
  $l\times L\times W=2.7\times30\times7.2$ mm$^3$. The intercontact
  distance $d=ab$ is 6.5 mm. Working temperature is 1.79 K. A dc
  current $I=19$ A is applied in the $y$ direction ($b_I=16.6$ G).
  The step amplitude is $2b_0=6$ G. The origin of time is taken at the
  very beginning of the increasing step. The signal $V_{ab}(t)$ is
  amplified and then analyzed by points through a Boxcar
  integrator.  The parasitic signal induced in the measuring
  loop (dashed line) has been calibrated at zero field, $B_0=0$. The
  pulse duration ($\sim10\;\mu$s) corresponds to the rising time of a
  step. From its shape, empirically well fitted by the difference of
  two exponentials, we obtain $db_e/dt$ in the form ({\ref{phi}}),
  then the shape $b_e(t)/b_0$ of the step, which is an instrumental
  constant. The amplitude of the parasitic signal provides an
  estimation of the effective surface of the measuring loop, $s=0.32$
  mm$^2$. Data reported in this figure, after substracting the
  parasitic signal, represent experimental values of the induced
  voltage $e(0,t)d$, at $B_0=3000$ G ($\bullet$). At 3000 G,
  $\sigma_f=1.76\times10^7\Omega^{-1}$m$^{-1}$, and $I_c=20.4$ A
  ($b_c=17.8$ G). From $b_c$ we infer that the diffusion is delayed
  and begins at time $t_d=1.6\mu$s. The full line represents the
  theoretical response $e(0,t)d$ such as calculated from
  Eq.~({\ref{champebis}}) by substituting $2l$ for $l$. The time
  constant $\tau_1\simeq0.1\tau=0.4\mu_0\sigma_fl^2\simeq65\;\mu$s is
  4 times longer than expected in a symmetric response.}
\end{figure}


\begin{thebibliography}{99}
\bibitem{Mathieu88} P.  Mathieu and Y. Simon, Europhys. Lett. {\bf 5}, 67 (1988). 
\bibitem{Hocquet92} T. Hocquet, P.  Mathieu and Y. Simon, Phys. Rev. B {\bf 46}, 1061 (1992).
\bibitem{Vasseur97} H. Vasseur, P. Mathieu, B. Pla\c{c}ais and Y. Simon, Physica C {\bf 279}, 103 (1997)
\bibitem{Campbell69} A.M. Campbell, J. Phys. C {\bf 2}, 1492 (1969).
\bibitem{Klundert78} L.J.M. van de Klundert, E.A. Gijsbertse and  H.P. van der Braak, Physica B {\bf 94}, 
41 (1978); E.A. Gijsbertse, M. Caspari and L.J.M. van de Klundert, Cryogenics {\bf 21}, 299 (1981). 
\bibitem{Abulafia95} Y. Abulafia, A. Shaulov, Y. Wolfus, R. Prozorov, L. Burlachkov, Y. Yeshurun, D. Majer, E. Zeldov and V.M. Vinokur, Phys. Rev. Lett. {\bf 75}, 2404 (1995).
\bibitem{Bean62} C.P. Bean, Phys. Rev. Letters {\bf 8}, 250 (1962).
\bibitem{Coffey92} M.W. Coffey and J.R. Clem, Phys. Rev. B {\bf 46}, 11757 (1992). 
\bibitem{Campbell72} A.M. Campbell and J.E. Evetts, Adv. Phys. B {\bf 21}, 199 (1972).
\bibitem{Josephson66} B.D. Josephson, Phys. Rev. {\bf 152}, 211 (1966).
\bibitem{Blatter94} G. Blatter, M.V. Feigel'man, V.B. Geshkenbein, A.I. Larkin and V.M. Vinokur, Rev. Mod. Phys. {\bf 66}, 1125 (1994).
\bibitem{Mathieu93} P. Mathieu, B. Pla\c{c}ais and Y. Simon, Phys. Rev. B {\bf 48}, 7376 (1993). 
\bibitem{Placais94} B.  Pla\c{c}ais, P.  Mathieu and Y. Simon, Phys. Rev. B {\bf 49}, 15813 (1994).
\bibitem{LuetkeEntrup97} N. L\"utke-Entrup, B. Pla\c{c}ais, P. Mathieu and Y. Simon, Phys. Rev. Lett. {\bf 79}, 2538 (1997).
\bibitem{Thorel72b} P. Thorel, R. Kahn, Y. Simon and D. Cribier, J. Phys. (Paris) {\bf 34}, 447 (1972). 
\bibitem{Thorel73} P. Thorel, Y. Simon and A. Guetta, J. Low Temp. Phys. {\bf 11}, 333 (1973).
\bibitem{Joiner67} W.C.H. Joiner and G.E. Kuhl, Phys. Rev. {\bf 163}, 362 (1967).
\bibitem{Simon94} Y. Simon, B.  Pla\c{c}ais and P.  Mathieu, Phys. Rev. B {\bf 50}, 3503 (1994).
\bibitem{LuetkeEntrup98} N. L\"utke-Entrup, B. Pla\c{c}ais, P. Mathieu and Y. Simon, to appear in Physica B.
\bibitem{MaggioAprile97}I. Maggio-Aprile, C. Renner, A. Erb, E. Walker and O. Fischer, Nature {\bf 390}, 487 (1997).
\bibitem{Thorel72} P. Thorel, Thesis, Paris (1972).
\bibitem{Prozorov98} R. Prozorov, E.B. Sonin, E. Sheriff, A. Shaulov and 
Y. Yeshurun, \emph{cond-mat/9802270}.
\bibitem{Farrel69} D.E. Farrel, B.S. Chandrasekhar and H.V. Culpert, Phys. Rev. {\bf 177}, 694 (1969).
\bibitem{Kawashima78} T. Kawashima, T. Ezaki and K. Yamajuji, Japan J. Appl. Phys. {\bf 17}, 551 (1978).
\bibitem{Alais67} P. Alais and Y. Simon, Phys. Rev. {\bf 158}, 426 (1967).


\end{thebibliography}
\end{document}